\documentclass[slac_one]{revtex4}
\usepackage{graphicx}
\usepackage{fancyhdr}
\begin{document}

\title{Lightest Neutralino Mass in the MNSSM} 

%

\author{S. Hesselbach, G. Moortgat-Pick}
\affiliation{IPPP, Durham, DH1 3LE, UK}
\author{D. J. Miller, R. Nevzorov}
\affiliation{University of Glasgow, Glasgow, G12 8QQ, UK}
\author{M. Trusov}
\affiliation{ITEP, Moscow, 117218, Russia}

\begin{abstract}
We argue that the allowed range of the mass of the lightest
neutralino in the MNSSM is limited. We establish the theoretical
upper bound on the lightest neutralino mass and obtain an
approximate solution for this mass.
\end{abstract}

\maketitle

\thispagestyle{fancy}


\section{INTRODUCTION}
Recent observations indicate that $22\%-25\%$ of the energy
density of the Universe exists in the form of stable
non--baryonic, non--luminos (dark) matter. Supersymmetric (SUSY)
models provide a good candidate for the cold dark matter component
of the Universe. If R--parity is conserved, the lightest
neutralino is absolutely stable and can play the role of dark
matter. However the minimal supersymmetric standard model (MSSM)
suffers from the so-called $\mu$--problem. The MSSM superpotential
contains only one bilinear term $\mu (\hat{H}_1 \epsilon
\hat{H}_2)$. The parameter $\mu$ is expected to be of the order of
$M_{Pl}$ or GUT scale. At the same time the correct pattern of
electroweak (EW) symmetry breaking requires $\mu\sim 100-1000\,
\mbox{GeV}$. The $\mu$--problem can be solved within the minimal
non--minimal supersymmetric standard model (MNSSM). The
superpotential of the MNSSM can be written as
\begin{equation}\label{1}
W=\lambda \hat{S}(\hat{H}_d \epsilon \hat{H}_u)+\xi
\hat{S}+W_{MSSM}(\mu=0)\,.
\end{equation}
At the EW scale the singlet superfield $\hat{S}$ gets non-zero
vacuum expectation value (VEV) $\langle S \rangle=s/\sqrt{2}$ and
an effective $\mu$-term ($\mu_{eff}=\lambda s/\sqrt{2}$) is
generated. Thus in the MNSSM the $\mu$ problem is solved in a
similar way to the next--to--minimal supersymmetric standard model
(NMSSM), but without the accompanying problems of singlet tadpoles
or domain walls.

\section{UPPER BOUND ON THE LIGHTEST NEUTRALINO MASS}
The neutralino sector of the MNSSM is formed by the superpartners of
the neutral gauge bosons ($\tilde{W}_3,\,\tilde{B}$) and neutral
Higgsino fields ($\tilde{H}^0_d$,\,$\tilde{H}^0_u$,\,$\tilde{S}$).
In the field basis $(\tilde{B},\,\tilde{W}_3,\,\tilde{H}^0_d,\,\tilde{H}^0_u,\,\tilde{S})$
the neutralino mass matrix takes a form
\begin{equation}\label{2}
M_{\tilde{\chi}}= \left(
\begin{array}{ccccc}
M_1                  & 0                  & -M_Z s_W c_{\beta}   & M_Z s_W s_{\beta}  & 0
\\[2mm]
0                    & M_2                & M_Z c_W c_{\beta}    & -M_Z c_W s_{\beta} & 0
\\[2mm]
-M_Z s_W c_{\beta}   & M_Z c_W c_{\beta}  &  0                   & -\mu_{eff}         &
-\frac{\lambda v}{\sqrt{2}} s_{\beta} \\[2mm]
M_Z s_W s_{\beta}    & -M_Z c_W s_{\beta} & -\mu_{eff}           &  0                 &
-\frac{\lambda v}{\sqrt{2}} c_{\beta} \\[2mm]
0                    & 0                  & -\frac{\lambda
v}{\sqrt{2}} s_{\beta}    & -\frac{\lambda v}{\sqrt{2}} c_{\beta}
& 0
\end{array}
\right)\,,
\end{equation}
where $s_W=\sin\theta_W$, $c_W=\cos\theta_W$,
$s_{\beta}=\sin\beta$, and $c_{\beta}=\cos\beta$. Here
$\tan\beta=v_2/v_1$ and $v=\sqrt{v_1^2+v_2^2}=246\,\mbox{GeV}$,
where $v_1$ and $v_2$ are the VEVs of $H_d$ and $H_u$,
respectively. The neutralino spectrum in the MNSSM may be
parametrised in terms of
\begin{equation}\label{3}
\lambda\,,\qquad \mu_{eff}\,,\qquad \tan\beta\,, \qquad M_1\,,\qquad M_2\,.
\end{equation}
In the minimal SUGRA inspired SUSY GUT's $M_1\simeq 0.5 M_2$.
The validity of perturbation theory up to the GUT scale implies
that $\lambda(M_Z)\le 0.7$\,. When $\lambda$ is small the
non--observation of Higgs bosons at LEP rules out $\tan\beta\le
2.5$\,. LEP searches for SUSY particles also set lower bounds on
$|M_2|,\,|\mu_{eff}|>90-100\,\mbox{GeV}$\,.

In order to find theoretical bounds on the neutralino masses
$m_{\chi_i^0}$ it is convenient to consider the matrix
$M_{\tilde{\chi}^0} M^{\dagger}_{\tilde{\chi}^0}$. In the basis
$\left(\tilde{B},\tilde{W}_3,-\tilde{H}^0_d
s_{\beta}+\tilde{H}^0_u c_{\beta}, \tilde{H}^0_d
c_{\beta}+\tilde{H}^0_u s_{\beta}, \tilde{S}\right)$ the
bottom-right $2\times 2$ block of $M_{\tilde{\chi}^0}
M^{\dagger}_{\tilde{\chi}^0}$ takes the form
\begin{equation}\label{4}
\left(
\begin{array}{cc}
|\mu_{eff}|^2+\sigma^2 \qquad\qquad           & \nu^{*}\mu_{eff} \\[1mm]
\nu\mu^{*}_{eff}                              & |\nu|^2
\end{array}
\right),
\end{equation}
where $\sigma^2=M_Z^2\cos^2 2\beta+|\nu|^2\sin^2 2\beta$,
$\nu=\lambda v/\sqrt{2}$. Since the minimal eigenvalue of any
hermitian matrix is less than its smallest diagonal element at
least one neutralino in the MNSSM is always light, i.e.
$|m_{\chi^0_1}|\le |\nu|$. Therefore in contrast with the MSSM the
lightest neutralino in the MNSSM remains light even when the SUSY
breaking scale tends to infinity. The mass of the lightest
neutralino must be also smaller than the minimal eigenvalue $\mu_0^2$
of the submatrix (\ref{4}), i.e.
\begin{equation}\label{5}
|m_{\chi^0_1}|^2\le \mu_0^2=
\frac{1}{2}\biggl[|\mu_{eff}|^2+\sigma^2+|\nu|^2-
\sqrt{\biggl(|\mu_{eff}|^2+\sigma^2+|\nu|^2\biggr)^2-4|\nu|^2\sigma^2}\biggr]\,.
\end{equation}
The value of $\mu_0$ decreases with increasing $|\mu_{eff}|$.
Taking into account the restrictions on $|\mu_{eff}|$ and
$\lambda(M_Z)$ we find that $|m_{\chi^0_1}|\le 80-85\,\mbox{GeV}$
\cite{Hesselbach:2007te}. The obtained theoretical restriction on
$|m_{\chi^0_1}|$ allows to discriminate the MNSSM from other SUSY
models.

\section{APPROXIMATE SOLUTION}
The mass of the lightest neutralino can be computed numerically by solving
the characteristic equation
\begin{equation}\label{7}
\begin{array}{c}
\mbox{det}\left(M_{\tilde{\chi}^0}-\kappa I\right)=
\biggl(M_1M_2-(M_1+M_2)\kappa+\kappa^2\biggr)
\biggl(\kappa^3-(\mu_{eff}^2+\nu^2)\kappa+\nu^2\mu_{eff}\sin
2\beta\biggr)\\[2mm]
+M_Z^2\biggl(\tilde{M}-\kappa\biggr)\biggl(\kappa^2+\mu_{eff}\sin
2\beta \kappa- \nu^2\biggr)=0\,,
\end{array}
\end{equation}
where $\tilde{M}=M_1 c_W^2 + M_2 s^2_W$. Because in the MNSSM
$|m_{\chi^0_1}|$ is normally much smaller than the masses of the other
neutralinos one can ignore $\kappa^3$, $\kappa^4$ and $\kappa^5$
terms in the characteristic equation so that it reduces to
\begin{equation}\label{8}
\kappa^2-B\kappa+C=0,
\end{equation}
where
$$
B=\frac{M_1 M_2}{M_1+M_2}+\biggl(\frac{\nu^2}{\mu_{eff}^2+\nu^2}-
\frac{M_Z^2}{\mu_{eff}^2+\nu^2}\frac{\tilde{M}}{M_1+M_2}\biggr)
\mu_{eff}\sin
2\beta-\frac{M_Z^2\nu^2}{(M_1+M_2)(\mu_{eff}^2+\nu^2)}\,,
$$
$$
C=\frac{\nu^2}{\mu_{eff}^2+\nu^2}\biggl(\frac{M_1M_2}{M_1+M_2}\mu_{eff}\sin
2\beta- \frac{\tilde{M}}{M_1+M_2}M_Z^2\biggr)\,.
$$
Then the mass of the lightest neutralino can be approximated by \cite{Hesselbach:2007te}
\begin{equation}\label{10}
|m_{\chi^0_1}|=\mbox{Min}\left\{\frac{1}{2}\biggl|B-\sqrt{B^2-4C}\biggr|,\,
\frac{1}{2}\biggl|B+\sqrt{B^2-4C}\biggr|\right\}.
\end{equation}

\begin{figure*}[t]
\centering
~\hspace*{-9cm}{$|m_{\chi_1^0}|$}\\
\includegraphics[width=100mm]{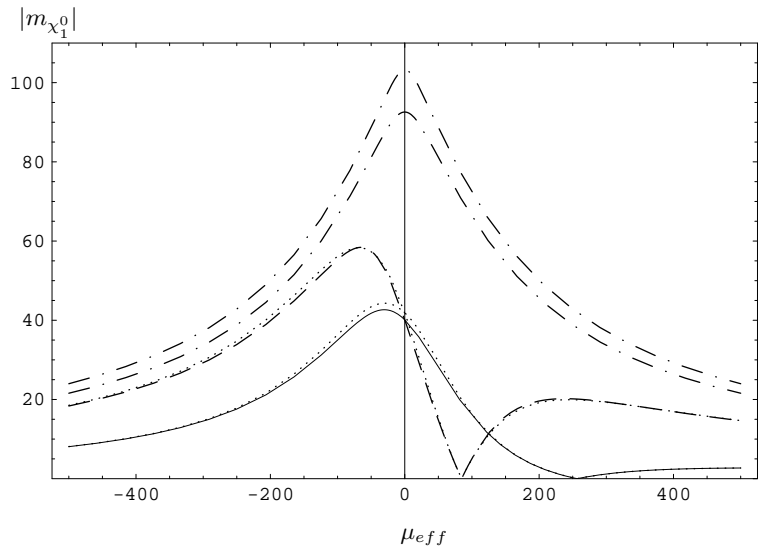}\\
~\hspace*{1cm}{$\mu_{eff}$}\\
\caption{Lightest neutralino mass versus $\mu_{eff}$ for
$\lambda=0.7$, $M_1=0.5 M_2$, $M_2=200\,\mbox{GeV}$. Solid and
dashed lines correspond to $\tan\beta=10$ and $3$. Upper and lower
dashed--dotted lines represent the theoretical restriction,
Eq.~(\ref{5}), on $|m_{\chi^0_1}|$ for $\tan\beta=3$ and $10$. Dotted lines
correspond to the approximate solution, Eq.~(\ref{10}).} \label{neutr-mu.eps}
\end{figure*}

\begin{figure*}[t]
\centering
~\hspace*{-9cm}{$|m_{\chi_1^0}|$}\\
\includegraphics[width=100mm]{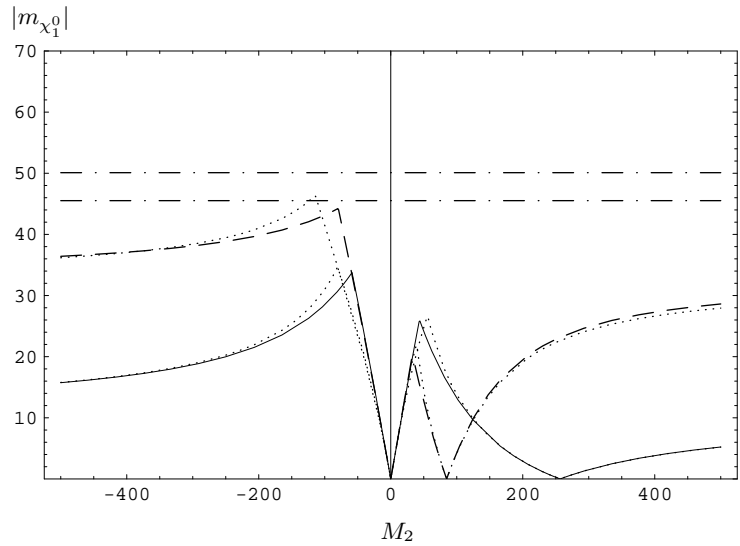}\\
~\hspace*{0.5cm}{$M_2$}\\
\caption{Lightest neutralino mass as a function of $M_2$ for
$\lambda=0.7$, $M_1=0.5 M_2$, $\mu_{eff}=200\,\mbox{GeV}$. The
notations are the same as in Fig. 1.} \label{neutr-gaugino.eps}
\end{figure*}

As follows from Fig.~1--2 the approximate solution describes
the numerical one with relatively high accuracy. With increasing $\mu_{eff}$ and
$M_{1,2}$ the lightest neutralino mass decreases. If either
$|\mu_{eff}|$ or $M_{1,2}\gg M_Z$ then
\begin{equation}\label{11}
|m_{\chi^0_1}|\simeq\frac{|\mu_{eff}|\nu^2\sin
2\beta}{\mu^2_{eff}+\nu^2}\,.
\end{equation}
According to Eq.(\ref{11}) $m_{\chi^0_1}$ is inversely
proportional to $\mu_{eff}$. It decreases when $\tan\beta$ grows
and vanishes when $\lambda$ tends to zero. When $\lambda\to 0$ the
mass of the lightest neutralino is proportional to $\lambda^2$.
When $|m_{\chi^0_1}|\ll M_Z$ the lightest neutralino is
predominantly singlino that makes its direct observation at future
colliders rather challenging.

\begin{acknowledgments}
RN would like to thank A. Pilaftsis and M. Vysotsky for fruitful
discussions.

RN acknowledges support from the SHEFC grant HR03020 SUPA 36878.
\end{acknowledgments}

\end{document}